\documentclass[12pt]{article}
\usepackage{natbib}
\usepackage{graphicx}
\usepackage{epsfig}
\usepackage{amssymb}

\begin{document}

\baselineskip24pt

\centerline{\bf Hungaria Asteroid Family as the Source of Aubrite Meteorites} 

\bigskip
\centerline{Matija \' Cuk$^1$, Brett J. Gladman$^2$, David Nesvorn\' y$^3$}

\bigskip

\centerline{$^1$Carl Sagan Center, SETI Institute}
\centerline{189 North Bernardo Avenue, Mountain View, CA 94043} 

\centerline{$^2$Department of Physics and Astronomy, University of British Columbia}
\centerline{6224 Agricultural Road, Vancouver, BC V6T 1Z1, Canada}

\centerline{$^3$Southwest Research Institute}
\centerline{1050 Walnut St, Suite 400, Boulder, CO 80302}

\bigskip

\centerline{E-mail: cuk@seti.org}

\vspace{24pt}
\centerline{Re-submitted to Icarus}
\centerline{May 23$^{\rm rd}$ 2014.}

\vspace{24pt}

\centerline{Manuscript Pages: 31}

\centerline{Figures: 6}

\centerline{Tables: 2}
\newpage

Proposed Running Head: Aubrites from Hungaria Family

\vspace{48pt}

Editorial Correspondence to:

Matija \' Cuk

Carl Sagan Center

SETI Institute

189 N Bernardo Ave

Mountain View, CA 94043

Phone: 650-810-0210

Fax: 650-

E-mail: mcuk@seti.org

\newpage

\noindent {ABSTRACT: The Hungaria asteroids are interior to the main asteroid belt, with semimajor axes between 1.8 and 2 AU, low eccentricities and inclinations of 16-35 degrees. Small asteroids in the Hungaria region are dominated by a collisional family associated with (434) Hungaria. The dominant spectral type of the Hungaria group is the E or X-type \citep{war09}, mostly due to the E-type composition of Hungaria and its genetic family. It is widely believed the E-type asteroids are related to the aubrite meteorites, also known as enstatite achondrites \citep{gaf92}. Here we explore the hypothesis that aubrites originate in the Hungaria family. In order to test this connection, we compare model Cosmic Ray Exposure ages from orbital integrations of model meteoroids with those of aubrites. We show that long CRE ages of aubrites (longest among stony meteorite groups) reflect the delivery route of meteoroids from Hungarias to Earth being different than those from main-belt asteroids. We find that the meteoroids from Hungarias predominantly reach Earth by Yarkovsky-drifting across the orbit of Mars, with no assistance from orbital resonances. We conclude that the CRE ages of aubrites are fully consistent with a dominant source at the inner boundary of the Hungaria family at 1.7~AU. From here, meteoroids reach Earth through the Mars-crossing region, with relatively quick delivery times favored due to collisions (with Hungarias and the inner main-belt objects). We find that, after Vesta, (434) Hungaria is the best candidate for an asteroidal source of an achondrite group.}

Key words: asteroids; asteroids, dynamics; meteorites; planetary dynamics; celestial mechanics.

\newpage 

\section{Introduction}

Hungarias are a dynamical group of asteroids interior to the asteroid belt but exterior to the orbit of Mars (in the 1.8-2 AU range). Most stable Hungarias have high inclinations (16-35$^{\circ}$) and low eccentricities ($<0.1$). Hungarias are bordered in inclination by multiple secular resonances, separated from the main asteroid belt by the $\nu_6$ secular resonance and the 4:1 mean-motion resonance with Jupiter, with a less well-defined inner boundary forced by close encounters with Mars \citep{war09, mil10}.

Unlike the main asteroid belt, Hungarias are not stable over the age of the Solar System, but are escaping into the Mars-crossing region with a half-lives in the 500-1000 Myr range \citep{mil10, mce10}. In the early Solar System the proto-Hungarias were shown to be orders of magnitude more numerous, and are proposed to be the main source of the Late Heavy Bombardment \citep{cuk12, bot12}. Hungarias may be depleted remnants of primordial Mars-crossers \citep{cuk12}, or survivors from the extinct innermost part of main asteroid belt perturbed by late planetary migration \citep{bot12}. Hungarias may therefore contain survivors from ancient collisional families which are now almost completely extinct via this large-scale dynamical depletion. This may help resolve some paradoxes in the meteorite-asteroid connection, like the lack of an extant suitable candidate for the mesosiderite parent body \citep{cuk12}.

While they inhabit a single island of relative dynamical stability, Hungarias are not all compositionally uniform, with most common asteroid types being S and E \citep{car01, war09}. A significant fraction of Hungarias belong to the Hungaria Genetic Family (HGF), centered on (434) Hungaria \citep{wil92, lem94, war09}\footnote{In this paper, "Hungaria" and "Hungarias" refer to all relatively stable asteroids in the Hungaria zone 1.8-2~AU, regardless of the spectroscopic type; we use "(434) Hungaria" for their largest member and "HGF" for its collisional family.}. This family consists of otherwise rare E-type asteroids, which are a proposed source of aubrite meteorites \citep{gaf92}. Apart from the HGF, Hungarias may contain another collisional family at slightly higher inclinations, comprised of S-type asteroids \citep{wil92, mil10}. As this group is relatively spread out and S-types are common in the inner asteroid belt, it is not yet established that this is a real genetic family, and we will not address it here.  

Aubrites, or enstatite achondrites, are the second most common group of achondrites, after the HED complex \citep{lor03}. Most aubrites are thought to originate from a single parent body, a large differentiated planetesimal that was disrupted soon after the formation of the Solar System. The immediate precursor bodies of aubrites can therefore be much smaller than the original progenitor, which was likely lost during the turbulent early history of the Solar System. The anomalous Shallowater aubrite, however, likely derives from the separate parent body from most aubrites \citep{kei89b}.

E-type asteroids are widely seen as likely source of aubrites, as their spectra are consistent with high-albedo iron-poor minerals like enstatite that dominate the aubrites \citep{zel75, zel77, gaf92, clo93, pie94, for01, bur02, kel02, cla04, for08}. Three spectral sub-types of E-type asteroids have been identified in the literature \citep{cla04, for08}. E[I] asteroids have featureless spectra characteristic of aubritic pyroxene plus feldspar assemblage; E[II] presents the strong absorption at 0.49 micron due to sulfide such oldhamite and occasionally at 0.90-0.96 micron; E[III] have absorption bands in the 0.9 and 1.8 micron region, indicating a silicate mineralogy higher in iron than the mineral enstatite. These subgroups likely possess distinct surface mineralogies, and more than one sub-type is identified within the HGF \citep{for11}. While the surface of (434) Hungaria itself may contain heterogenities \citep{for08}, most recent data indicate that Hungaria exhibits 0.49 micron band typical of the E[II] subtype \citep{for11}. Some of E-types (notably 44 Nysa) appear to have hydration feature in the infra-red (3-micron band), which is not consistent with aubrite composition \citep{riv95}. However, alternative explanations for the presence of this band have also been put forward, in which case at least a partial aubrite composition cannot be excluded \citep{gaf02}. \citet{cla04} find that E-types within the HGF typically match aubrite visible and near-infrared spectra significantly better than non-Hungaria E-type asteroids (like 44 Nysa and 64 Angelina), as spectra of both E(II) asteroids and aubrites tend to be featureless. However, some individual aubrites can be well matched by main belt E-type asteroids \citep{for08}. 

Apart from spectroscopy, aubrites' long Cosmic Ray Exposure (CRE) ages indicate that they may be delivered through a distinct dynamical route from most stony meteorites \citep{lor03, eug06, kei10}. Aubrites have median cosmic ray exposure ages of 50 Myr (Table \ref{aubrite}), significantly longer than those of other stony meteorites. Median CRE ages for H, L, and LL ordinary chondrites are 7~Myr, 20~Myr and 15~Myr, respectively\citep{mar92, eug06}, and about 20~Myr for HED meteorites \citep{eug06}. Differences in CRE ages of stony and iron meteorites imply that collisional destruction plays the major role in limiting the lifetime of small meteoroids \citep{mor98}. There is however no reason to think that highly brecciated aubrites are stronger than other stones, so their long CRE ages would more likely be a consequence of their orbital history \citep{her03}. A unique orbital history could be consistent with an origin in a distinct dynamical group such as Hungarias, rather than the main asteroid belt, where the overwhelming majority of meteorites originate.

Since Hungarias are interior to strong Jovian resonances but adjacent to Mars, the primary route for their escape (and potential Earth impact) is through Mars-crossing orbits. Median aubrite CRE ages of 50 Myr are comparable to typical Mars-crosser dynamical lifetimes of 80 Myr \citep{cuk12}, supporting a possible Hungaria-aubrite connection. The Mars-crossing region is also depleted in meteoroidal debris, allowing for much longer collisional lifetimes of stony meteoroids on low $e$-orbits, than is possible in the main asteroid belt (or Hungarias). 

In contrast to Mars-crossers, Earth-crossers have shorter dynamical lifetimes, with an average of 15 Myr \citep{gla00}. This short lifetime of Near-Earth Asteroids (NEAs) was not recognized until the late 1990s when direct integrations replaced \" Opik-type calculations. Previous suggestions of aubrite origin on (3101) Eger \citep{gaf92}, or any other NEA, are thus not consistent with the modern understanding of NEA dynamical timescales.

In this paper, we will study the Hungaria-aubrite connection using numerical simulation of the motion of test particles originating among Hungarias and subject to planetary perturbations. These simulated meteoroids are also subject to two non-gravitational effects: the Yarkovsky drift \citep{far99} and collisional removal. In the following sections we will explore how different starting points of meteoroids and models of these two processes affect the postulated Hungaria-aubrite connection. In particular, we will attempt to find a self-consistent model of aubrite delivery which could explain the CRE ages of known aubrites.

\bigskip

\section{Dynamics of meteoroids launched from (434) Hungaria}

\bigskip
The first numerical experiment aimed at testing the Hungaria-aubrite connection involved the integration of 4800 test-particles using SWIFT-rmvsy integrator \citep{bro06}. This integrator is based on the standard SWIFT-rmvs3 mixed-variable symplectic algorithm which is capable of integrating close approaches between planets and small bodies \citep{lev94}, and is further modified by incorporating the Yarkovsky thermal recoil force on small bodies \citep{rub95, far99, bot06}. This force depends on the size and spin properties of the body; we assumed uniform sizes with radii $R=1$~m, obliquities of 45$^{\circ}$ and spin period of about one minute with half of the sample being direct and half retrograde rotators. Bodies with radius $R=1$~m were used as they are likely the largest (and therefore longest lived) meteoroids that are fully penetrated by cosmic rays. For SWIFT-rmvsy input, we used specific heat of 1000 J~kg$^{-1}$~K$^{-1}$, and thermal conductivity of 0.1~W~m$^{-1}$~K$^{-1}$ \citep[previously found appropriate for stony meteoroids by ][]{bot06}, with uniform density of 3000~kg~m$^{-3}$, implying solid-rock composition. In this regime the seasonal Yarkovsky effect dominates and all bodies migrate inward, toward the Sun \citep{bot06}. The test particles were affected by the gravity of the eight planets, but not each other. All particles were initially placed close to the orbit of (434) Hungaria, with orbital velocities spread within 1\% of that of Hungaria. To do this, we used a regular grid to assign small kicks to v and z components of the particles' velocities (in the ecliptic coordinate system). Particles were followed until they collided with the Sun or planets, or were ejected from the system.
 
\begin{figure*}[h]
\includegraphics[scale=.7, angle=270]{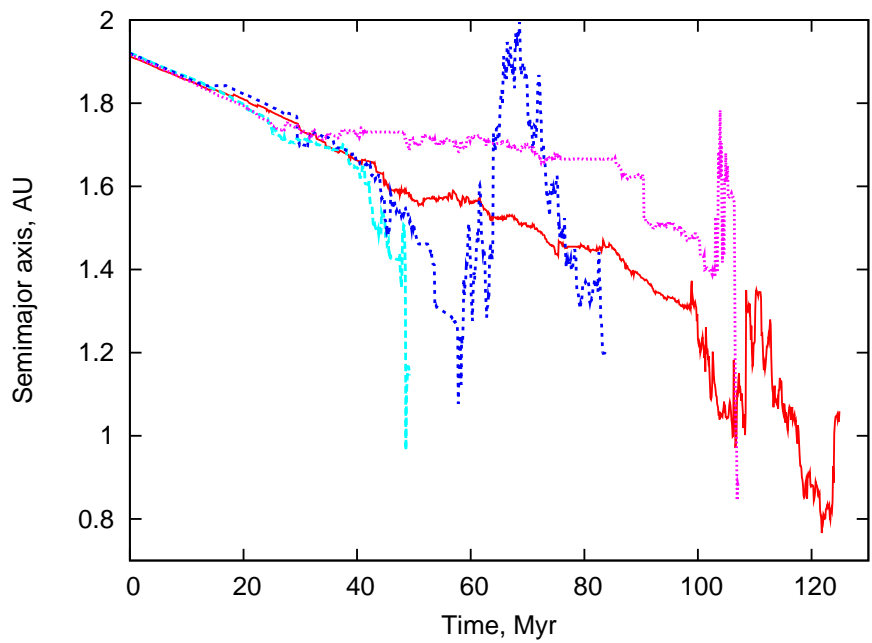}  
\caption{Exaples of semimajor evolution of four model earth-impacting meteoroids starting at 1.94~AU and evolving under the Yarkovsky effect and planetary perturbations. All four particles are assumed to be retrograde rotators.} 
\label{foura}
\end{figure*}

\begin{figure*}[h]
\includegraphics[scale=.7, angle=270]{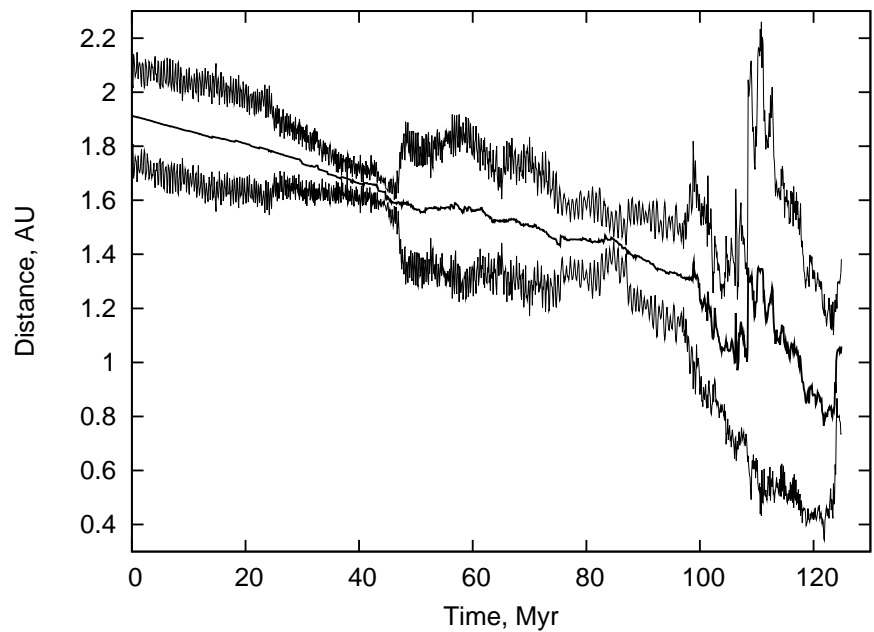}  
\caption{Evolution of aphelion, perihelion and semimajor axis (top, bottom and middle lines, respectively) for the longest-lived particle featured in Fig. \ref{foura}}. 
\label{qaq}
\end{figure*}

We find that direct rotators migrate toward the Sun at 0.002~AU~Myr$^{-1}$, while retrograde ones spiral in at 0.005~AU~Myr$^{-1}$, in rough agreement with Fig. 2a in \citet{bot06}, who used somewhat lower heat capacity and meteoroid density. Fig. \ref{foura} plots semimajor axis evolution of four particles that eventually hit Earth (all four were retrograde rotators). It is clear that the evolution is completely determined by the Yarkovsky drift (apart from crossings of minor resonances) until the meteoroids reach about 1.7~AU, where scattering by Mars becomes important. Once Mars-crossing, particles take as little as 10 Myr or as long as 100 Myr to become Earth-crossing. Before the onset of Earth encounters, eccentricities tend to be low or moderate, but once Earth-crossing occurs, eccentricities become typical of NEAs and the meteoroids' aphelia are often in the inner asteroid belt. Fig. \ref{qaq} shows the perihelion and aphelion evolution of one of the particles from Fig. \ref{foura}. The particle in question acquires a large eccentricity only after the onset of Earth encounters at 100 Myr. While some particles spend lots of time on asteroid-belt-crossing orbits, this pattern of large periods spent on relatively low-$e$ Mars-crossing orbits is more typical of the simulated meteoroids. This behavior implies that the meteoroids can accumulate significant cosmic ray exposures while being relatively safe from collisions with the main asteroid belt (or Hungaria group) objects.

\begin{figure*}[h]
\includegraphics[scale=.6, angle=270]{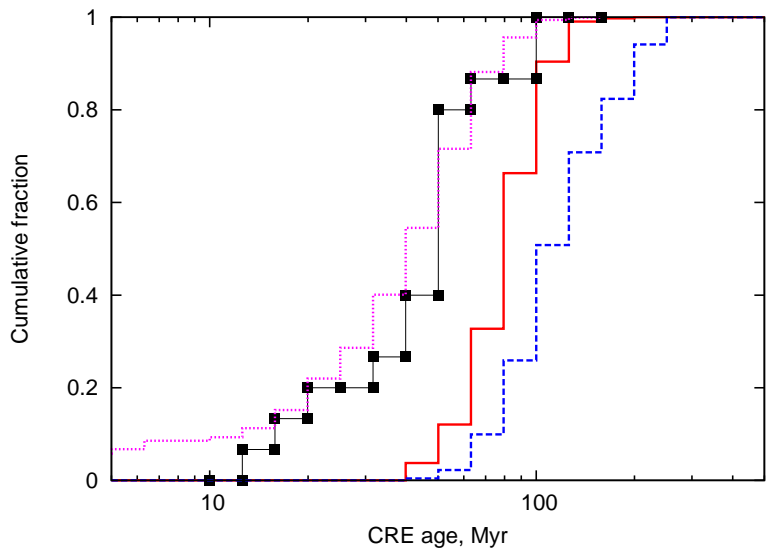}  
\caption{Cumulative distribution of CRE ages for 15 real aubrites (thin solid line with squares) and model distributions based on test particles released from (434) Hungaria. Dashed line plots the raw delivery times for meteoroids impacting Earth, while the solid line includes colisional removal beyond 1.7 AU with a timescale of 10 Myr. Dotted line shows the simulated CRE ages assuming that meteoroids' CRE clocks are reset first time they have $a<1.7$~AU.} 
\label{cumul1}
\end{figure*}

In total, 493 particles hit Earth, and the distribution of their impact times is plotted in Fig. \ref{cumul1} (dashed line). It is clear that these times are too long when compared to CRE ages of aubrites \citep[black line with square points, Table \ref{aubrite};][]{lor03}. This is normal when neglecting collisional disruption; the same is true for martian meteorites \citep{gla97}. The first necessary correction is to account for collisional loss during interplanetary transfer. Stony meteorites have median CRE ages of 20~Myr or less because of limited collisional lifetime of 1 m radius stony meteoroids. Since some of the dynamical delivery mechanisms (like jovian resonances) are faster than that, it is now thought that much of the CRE age is accumulated in the main belt during slow Yarkovsky drift into resonances \citep{mor98, bot06}.  Iron meteorites, being more resistant to disruption, have longer CRE ages, most of which is presumably accumulated on stable orbits in the asteroid belt. While our particles spend most of their time in the inner Solar System, they often traverse regions also occupied by main belt asteroids and associated debris, leading to some collisional evolution. 

We weigh the contribution of impacts with the function $p=exp(-t_c/\tau)$, where $t_c$ is the amount of time particle spends outside $r_0$=1.7~AU (the outer edge of the zone cleared by Mars), and $\tau$=10~Myr is corresponding assumed collisional lifetime.\footnote{A collisional lifetime of about $10^7$ years for 1~m bodies is well-established in the literature \citep{wet85, far98, mor98}. In Section 4 we explore the dependence of our results on the meteoroid collisional lifetime.} For each output interval, the fraction of the orbital period spent outside $r_0$is found by first finding the eccentric anomaly $E_0$ for which $r_0=a(1-e \cos{E_0})$. $E_0$ is then converted into the mean anomaly $M_0=E_0 - e \cos{E_0}$ \citep[we assume $0 < E_0 < \pi$;][]{md99}. The fraction of the orbit spent outside $r_0$ is then $M_0/\pi$. The integrated collisional time is then just $t_c=\Sigma (M_0/\pi) \Delta t$, summed over all the output intervals. Here we have ignored all details about radial and latitudinal distribution of meteoroids in the Hungarias and the inner main belt. This was done to avoid introducing parameters that are completely unconstrained by data, and keep the model as simple as possible. In Section 4 we will discuss in more detail how sensitive our results are to our choice of $\tau$.

This weighted distribution is plotted in Fig. \ref{cumul1} using a thick solid line. While collisional removal lead to shorter model CRE ages of surviving meteoroids, the model is still unable to match the observation. We can conclude that aubrites did not originate directly on (434) Hungaria or anywhere deep within the HGF. This is not surprising because the collective cross-section of HGF members is much larger than that of (434) Hungaria; most meteoroids should therefore be launched from smaller family members. In the next Section we will discuss likely source regions at the edge of HGF for meteoroids that are likely to survive the trip to Earth.

\bigskip

\section{Dynamics of meteoroid precursor bodies}

\bigskip

A very important discrepancy between the first integration and the aubrite data is that we do not produce any Earth impacts within the first 40~Myr. This is independent of our collisional removal scheme, as we simply have no meteoroids that make it from (434) Hungaria to Earth in that time. The fact that we have aubrites with CRE ages as low as 10-20~Myr means that they could not have been released at low relative speeds from their parent bodies deep within the stable zone at 1.94~AU where (434) Hungaria resides, but rather from somewhere closer to the Mars-crossing region. This should not be surprising, because modeling of the Yarkovsky-driven dispersal of the family \citep{war09} shows that one expects $H>18$ ($D<$1 km) fragments to have dispersed today down to as low as $a=1.8$~AU, with smaller-yet meteorite parent bodies potentially even closer. Bodies with 10~m$<D<$100~m can drift as fast as 2m meteoroids \citep{bot06}, but also have longer collisional lifetimes, which makes them better candidates for transporting aubrite material down to 1.7 AU. A short residence in the relatively populated Hungaria region \citep[which is also exposed to bombardment by inner-belt asteroids, such as Floras, at their perihelia;][]{war09} should strongly favor the survival of meteoroids released as close to the Mars-crossing region as it is realistic.

\begin{figure*}[h]
\includegraphics[scale=.6, angle=270]{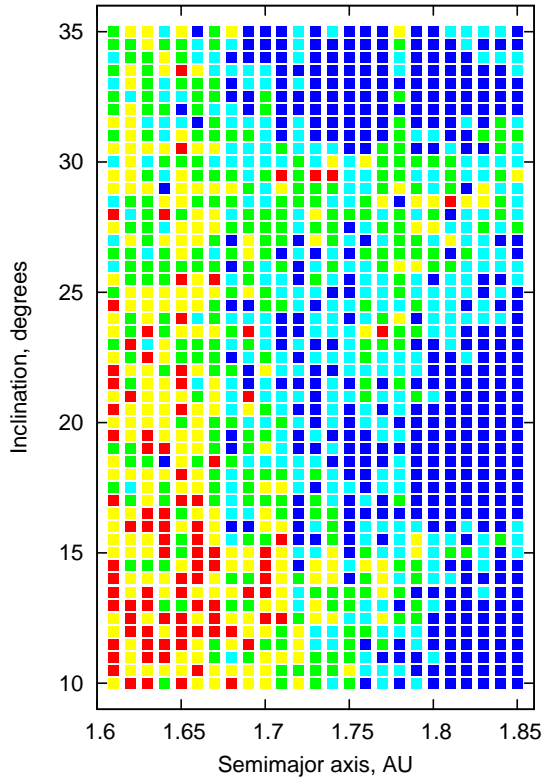}
\caption{Stability map of the inner boundary of the Hungaria region. Points plot initial conditions for $e=0$ test particles, with the colors indicating the time $t_{\Delta}$before a particle's semimajor axis diverged by 0.05 AU from the initial value. Red: $t_{\Delta} < 3$~Myr; yellow: 3~Myr $< t_{\Delta} <$10~Myr; green: 10~Myr $<t_{\Delta}<$ 30 Myr; cyan: 30~Myr $<t_{\Delta}<$ 100~Myr; blue: $t_{\Delta}>$ 100~Myr.}
\label{stab}
\end{figure*}

In order to explore the distribution of likely meteoroid parent bodies in the HGF, we used a purely-gravitational SWIFT-rmvs4 integrator to study the stability of test particles at the inner edge of the Hungaria region against perturbations (primarily close encounters with Mars). We integrated a grid of 1275 particles in semimajor axis and inclination (with initially circular orbits) over 300 Myr. The results of this simulation are plotted in Fig \ref{stab}. Different colors in the semimajor axis-inclination space indicate orbits that are more stable (blue and cyan) versus those that are less stable (red and yellow). Stability is measured by the amount of time it took the particles to have their semimajor axes perturbed at least 0.05~AU from their initial conditions. 

It is clear that the orbits interior to about 1.7 AU tend to be variable on short timescales, while bodies outside 1.7~AU can have their orbits relatively unchanged for tens of millions of years. Note that the collisional lifetime of 10-100~m asteroids that we envision as the immediate precursor bodies of meteoroids is only a few tens of Myr \citep{bot06}. We conclude that 10-100~m parent bodies of meteoroids form a halo around the observed Hungaria family. The inner edge of this extended halo is at about 1.7~AU, where depletion through encounters with Mars is balanced by supply of HGF fragments which are drifting inward due to the Yarkovsky effect. This inner edge of Hungarias is the most likely source of meteorites: meteoroids launched from here can be perturbed by Mars away from the remaining Hungarias faster then the meteoroids can be destroyed in mutual collisions.  
  
A quick test of this estimate of the meteoroid launch location would be to start the "clock" of the CRE ages of the meteoroids released at (434) Hungaria only once they cross 1.7~AU. So we re-analyzed our first SWIFT-rmvsy simulations by starting the CRE and collisional clocks once the particle has $a <$~1.7~AU for the first time. The dotted line in Fig. \ref{cumul1} plots the result of this exercise. The overall distribution of CRE ages computed that way matches the data much better. This is just an indicative estimate and should not be taken as a final result due to some artificial assumptions that went into the calculation. For example, the tail of very short CRE ages is caused by a handful of Earth-impacting particles that became Mars-crossing and Earth-crossing without having $a <$~1.7~AU (except just before the impact). In reality, these meteoroids' long and slow evolution through the inner asteroid belt would make their survival against collisional destruction very unlikely. To explore the delivery or Hungaria-derived meteoroids from 1.7~AU more realistically, we need to run a separate simulation with initial locations at 1.7~AU.  

\bigskip

\section{Dynamics of meteoroids launched at 1.7 AU}

\bigskip

In order to produce a more realistic distribution of Earth impact ages, we re-ran the Yarkovsky simulation, with the particles now starting at 1.7~AU (with eccentricities and inclinations still similar to that of 434 Hungaria). All the simulation settings were identical to the first simulation (described in Section 2), with the exception of meteoroid radii, which were now set to 0.75~m. We chose to decrease the meteoroid radii to see if the Yarkovsky drift can be accelerated, but the drift rates in the simulation did not exhibit any significant change, indicating that the Yarkovsky effect is not sensitively dependent to size in the model we are using.

\begin{figure*}[h]
\includegraphics[scale=.6, angle=270]{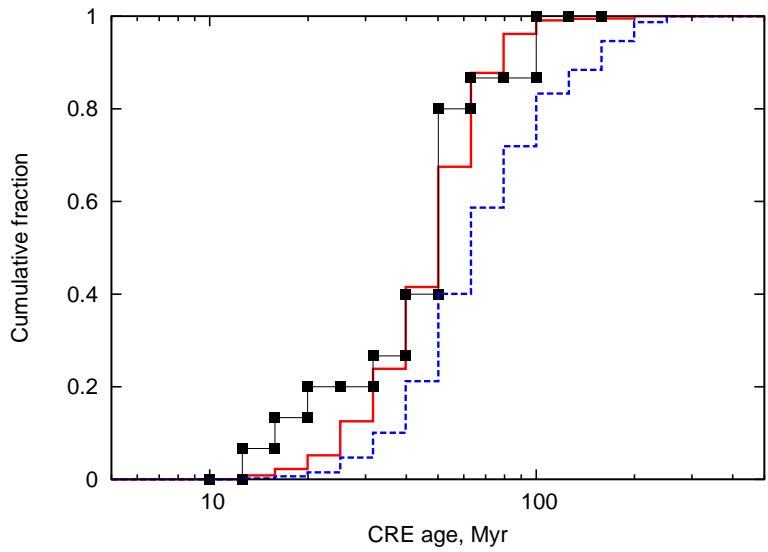}  
\caption{Cumulative distribution of CRE ages for 15 aubrites (thin solid line with squares) and model distributions based on test particles released at 1.7~AU (the inner edge of Hungaria region). Dashed line plots the raw delivery times for meteoroids impacting Earth, while the solid line includes colisional removal beyond 1.7 AU with a timescale of 10 Myr.}
\label{cumul2}
\end{figure*}

In Fig. \ref{cumul2}, the raw Earth impact times are plotted by a dashed line, and those corrected for collisional removal with a solid red line. Collisional removal was done in the same manner as in Fig. \ref{cumul1}, with the critical radius of 1.7~AU and collisional lifetime of $\tau=10$~Myr. It is evident that the collision-corrected population starting at 1.7~AU represents a good match to the available aubrite CRE data (solid line with square points).  Table \ref{table2} lists statistical comparisons between different model CRE ages in the two Yarkovsky simulations. Given that the number of "principal group" aubrites with known CRE ages is small (fifteen; Table \ref{aubrite}), statistical comparisons between the data and the models are not very constraining. Meteoroids starting at 1.94~AU are ruled out by a K-S test from being the aubrite precursors, as are meteoroids originating at 1.7~AU but experiencing no collisional evolution. However, a range of collisional evolution timescales (3--30~Myr) seems to match the data relatively well, so it is not possible to say what a precise collisional evolution timescale needs to be. Note that there is a two order of magnitude difference in the number of surviving bodies between models using $\tau=30$~Myr and $\tau=3$~Myr. While it is hard to estimate the absolute numbers of 1~m radius bodies among Hungarias, we think that relatively high frequency of aubrites among falls argues against the more rapid rate of comminution among Hungarias implied by $\tau=3~Myr$.

While the number of known aubrites with measured CRE ages is relatively small, the data appear to contain more bodies with short CRE ages than the model generates.\footnote{The Kolmogorov-Smirnov test does not require this, but this statistical test is known to be less sensitive to the tails of the distribution than to the center.} This is mostly due to the Aubres meteorite, with a CRE age of about 13~Myr. Our model using an origin at 1.7AU and $\tau=10$~Myr predicts that only 0.8\% of meteoroids should have CRE ages below 15~Myr. The chance of one of 15 known aubrites having such age is 12\%, according to the model. However, this estimate is based on only one particle in the second numerical simulation that impacted Earth this early, which has about 50\% probability of surviving collisional evolution on the way to Earth. Given the diversity of paths taken by our particles on the way to Earth (Fig. \ref{foura}), uncertainties on these numbers are enormous. Further numerical simulations involving orders of magnitude more particles could clarify the issue, but the comparison with data would still be frustrated by small numbers of known aubrite CRE ages. So while it is possible that there is a mismatch between data and the model at small CRE ages (perhaps indicating an additional delivery mechanism), there is currently no statistical support for requiring sources of aubrites other than Hungaria region.

Our collisional evolution model is very simplified and leaves out may details of the meteoroidal collisional cascade. We assume that the meteoroids are released at 1.7~AU as 1~m radius bodies, although meteoroids of that size are certainly released throughout the Hungaria region. While most of the bodies starting deeper among Hungarias are unlikely to survive collisions during the slow Yarkovsky-driven passage to the Mars-crossing region, such bodies would shift the model CRE ages to larger values. We also assumed that the multi-meter precursor bodies stop at 1.7~AU and none of them become Mars-crossers. This is certainly incorrect, and as most Mars-crossing orbits cross the inner asteroid belt, it is likely that many such parent bodies would be disrupted during their (on average) 80-Myr Mars-crossing phase. Ejection of ``fresh" meteoroids directly into Mars-crossers would shift the model CRE ages toward shorter ages. We decided against such additions as our study was mostly concerned with basic dynamics of Hungaria-derived meteoroids. Further studies could extend our results with more realistic collisional models, similar to how \citet{nes09} modeled the evolution of meteoroids from the Gefion family. In that study, \citet{nes09} made use of well-established dynamical results and incorporate them into a collisional model. Our understanding of Hungaria meteoroid dynamics is still behind that of main-belt derived bodies, that we think that the level in detail used in our present modeling is appropriate, given the limited amount of available CRE data.

\bigskip

\section{Conclusions}

In this work we studied the delivery of meteoroids from the Hungaria asteroid group using direct numerical integrations. We reached the following conclusions:

1. The published results on the surface spectral properties and composition of the E-type Hungaria Genetic Family (HGF, a subset of Hungaria asteroid group) support the HGF as the likely source of aubrite achondrite meteorite group (although some main-belt E-type asteroids cannot be excluded as sources of aubrites).

2. The long Cosmic-Ray Exposure (CRE) ages of aubrites support a distinct dynamical delivery mechanism relative to other stony meteorites.

3. Numerical simulations of meteoroid delivery from (434) Hungaria, which include the thermal Yarkovsky drift, give model CRE ages that are comparable, but still too long compared to those of known aubrites.

4. The relative stability of orbits surrounding the observed Hungaria group imply a more dispersed distribution of meteoroid immediate precursor bodies relative to the observable Hungaria-group asteroids. Collisional evolution strongly favors delivery of meteorites from bodies that are at the very boundary of the Mars-crossing region.

5. We find that meteorites launched from the inner boundary of the Hungaria zone, at 1.7~AU, would have CRE ages that match those of known aubrites. To obtain this match, we assume a collisional lifetime of typical meteoroid to be 3--30~Myr. 

6. We conclude that there is a good case for the collisional family centered on (434) Hungaria to be dominant source of aubrites (apart from anomalous ones, like Shallowater). 

While the HGF is not the overwhelming reservoir of E-type material \citep[unlike Vesta family for V-types;][]{bin93}, Hungarias are favorably placed to deliver meteoroids to Earth, compared to main-belt E-types. The relative rarity of E-types and the distinct CRE age signature of aubrites (arising from the unusual dynamics of Hungaria meteoroids) makes the connection between Hungarias and aubrites easier to establish than those for chondrite parent bodies \citep{nes09, bot10}.

\bigskip

\vspace{48pt}

{\centerline{\bf ACKNOWLEDGMENTS}}

\bigskip

M\' C is supported by NASA's Planetary Geology and Geophysics (PGG) program, award number NNX12AO41G. DN's work was also supported by the NASA PGG program. The numerical integrations were completed on the LeVerrier cluster at the University of British Columbia. The authors thank Joseph Burns and Douglas Hamilton for their help with this project. Two anonymous reviewers' suggestions greatly improved the paper.

\bibliographystyle{}

\newpage

\begin{table}
\begin{center}
\caption{CRE ages of 15 brecciated aubrites reported by \citet{lor03}. Anomalous Shallowater and Mt. Egerton aubrites were excluded as they may have originated on other parent bodies. Asterisk marks CRE ages that are an average between values for multiple fragments listed by \citet{lor03}.\label{aubrite}}
\bigskip
\begin{tabular}{|l|c|}
\hline\hline
Meteorite & CRE age (Myr)\\
\hline\hline
ALH~78113 & 21.2 \\
\hline
ALH~84007/84008/84011/84024 & 19.6* \\
\hline
Aubres & 12.6 \\
\hline
Bishopville & 52.0 \\
\hline
Bustee & 52.6 \\
\hline
Cumberland~Falls & 60.9 \\
\hline
EET~90033/90757 & 32.2 \\
\hline
Khor~Temiki & 53.9 \\
\hline
LEW~87007 & 53.9 \\
\hline
Mayo~Belwa & 117 \\
\hline
Norton~County & 111 \\
\hline
Pena~Blanca~Springs & 43.2 \\
\hline
Pasyanoe (light/dark) & 45.1* \\
\hline
QUE~97289/97348 & 40.6 \\
\hline
Y~793592 & 55.0 \\
\hline
\end{tabular}
\end{center}
\end{table}

\begin{table}
\begin{center}
\caption{Comparison between different numerical experiments, in which two orbital simulations of 4800 particles each have been analyzed using different collisional evolution assumptions. The columns list the initial distance in simulations (1.94 or 1.7~AU), timescale against collisional disruption $\tau$ (when beyond 1.7~AU), maximum difference of cumulative CRE age disribution from that of real aubrite CRE ages $D_{max}$, probability $p_{KS}$ of the model and the data being drawn from the same distribution (according to Kolmogorov-Smirnov test), number of Earth impacts $N_{Earth}$ (not an integer in collisonally-evolved models), fraction of Earth impacts $f_{Earth}$ and reference to the relevant Figures. Asterisk denotes the case where the particles were relased at 1.94~AU but the CRE clocks were started only at 1.7~AU (see text).\label{table2}}
\bigskip
\begin{tabular}{|c|c|c|c|c|c|l|}
\hline\hline
Initial distance & $\tau$ & $D_{max}$ & $p_{KS}$ & $N_{Earth}$ & $f_{Earth}$ & Remarks\\ 
\hline\hline
1.94~AU & infinite &	0.78 & $< 0.1$\% & 493 & 10.3\% & Fig. \ref{cumul1}, dashed\\
\hline
1.94~AU & 10~Myr & 0.68  & $< 0.1$\% & 5.9 & 0.1\% & Fig. \ref{cumul1}, solid\\
\hline
1.94 (1.7)~AU* & 10~ Myr & 0.15 & $> 20$\% & 80.5 & 1.7\% & Fig. \ref{cumul1}, dotted\\
\hline
1.7~AU & infinite & 0.4 & 1\% -- 2.5\% & 467 & 9.7\% & Fig. \ref{cumul2}, dashed\\ 
\hline
1.7~AU & 10~Myr & 0.15 & $> 20$\% & 60.7 & 1.3\% & Figs. \ref{cumul2}, \ref{cumul3}, solid\\
\hline
1.7~AU & 30~Myr	& 0.25 & $> 20$\% & 198.4 & 4.1\% & Fig. \ref{cumul3}, dashed\\
\hline
1.7~AU & 3~Myr	& 0.17 &  $> 20$\% & 4.2 & 0.1\% & Fig. \ref{cumul3}, dotted\\
\hline
\end{tabular}
\end{center}
\end{table}

\newpage
\newpage

\begin{figure*}[h]
\includegraphics[scale=.6, angle=270]{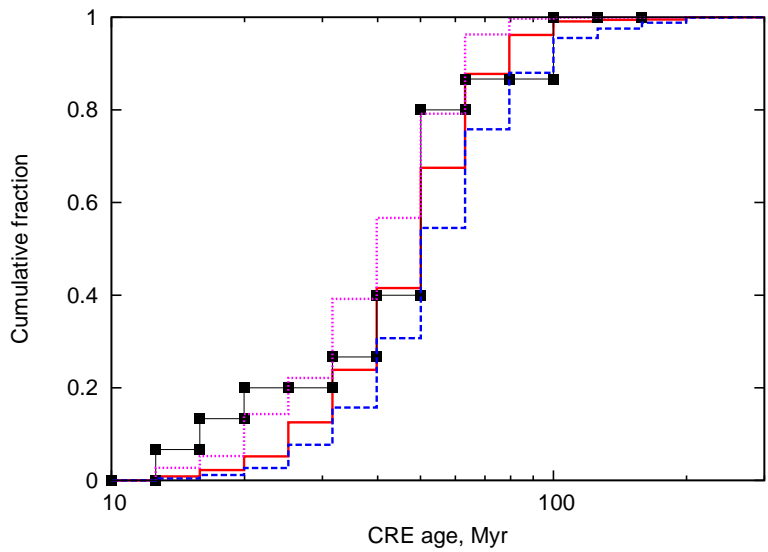}  
\caption{Cumulative distribution of CRE ages for 15 real aubrites (thin solid line with squares) and model distributions based the same orbital simulation as in Fig. \ref{cumul2}. Different lines plot simulated CRE age distributions, assuming collisonal lifetimes beyond 1.7~AU of 3~Myr (dotted), 10~Myr (solid) and 30~Myr (dashed line).}
\label{cumul3}
\end{figure*}


\end{document}